\documentclass{icrc29}
\usepackage{graphicx,amssymb,amsmath,times}
\begin{document}
\title[The Tunka Experiment: Towards ...]{The Tunka Experiment: Towards a 1-km$^2$ Cherenkov EAS Array in the Tunka Valley}
\author[N.M.Budnev et al.] {N.M.Budnev$^b$, D.V.Chernov$^a$, O.A.Gress$^b$, N.N.Kalmykov$^a$,
        V.A.Kozhin$^a$, E.E Korosteleva$^a$,
	\newauthor
	 L.A.Kuzmichev$^a$, B.K.Lubsandorzhiev$^c$, G.Navarra$^f$, 
	M.I.Panasyuk$^a$, L.V.Pankov$^b$,  
	\newauthor
	\framebox{Yu.V.Parfenov$^b$}, V.V.Prosin$^a$, V.S.Ptuskin$^d$, Yu.A.Semeney$^b$, A.V.Shirokov$^a$,
	 A.V.Skurikhin$^a$, 
	\newauthor
	C.Spiering$^e$, R.Wischnewski$^e$, I.V.Yashin$^a$\\  
	(a) Scobeltsyn Institute of Nuclear Physics MSU, Moscow, Russia,\\  
        (b) Institute of Applied Physics of ISU, Irkutsk, Russia,\\ 
	(c) Institute for Nuclear Research of RAN, Moscow, Russia,\\
        (d) IZMIRAN, Moscow, Russia,\\
	(e) DESY, Zeuthen, Germany,\\
	(f) Universita' Torino, Italy           
        }
\presenter{Presenter: L.A.Kuzmichev (kuz@dec1.sinp.msu.ru), \  
rus-kuzmichev-LA-abs1-he15-poster}

\maketitle

\begin{abstract}

The project of an EAS Cherenkov array in the Tunka valley/Siberia with an
area of about 1 km$^2$ is presented. The new array will have a ten times bigger area than the existing Tunka-24 array and will
permit a detailed study of the cosmic ray energy spectrum and the mass composition in the energy
range from 10$^{15}$ to 10$^{18}$ eV.

\end{abstract}

\section{Introduction}

Over the last decade, an impressive set of experimental data in the 
energy 10$^{14}$--10$^{16}$ has been collected(\cite{EASTOP}, \cite{Kaskade},
\cite{Space}, \cite{Tunka}). They yield a rather consistent picture and confirm the rise
in average mass of primary cosmic particles when passing the region of the 
"knee" at $3\cdot 10^{15}$ eV. The situation is dramatically different
at  higher energies. The energy range between 10$^{16}$ and 10$^{18}$ eV has been
covered by very few experiments. Our knowledge about mass composition above a few
10$^{16}$ eV is rudimentary. Energy spectra determined by different experiments 
differ significantly, mostly due to the difficulties in proper energy calibration. 
To the other hand the region above 10$^{16}$ eV is of crucial importance for understanding of
the origin and propagation of cosmic rays in the Galaxy. A careful investigation of this
region is mandatory and would answer the following questions:\\
-- Is there an "iron knee" above the classical knee at $3\cdot10^{15}$ eV? The
identification of an iron knee could provide a final understanding of this region.\\
-- What is the mass composition above a possible iron knee? Is this region 
dominated by the sources different to supernova remnants?\\
-- What is the nature of the observed "second knee" at $3-5\cdot 10^{17}$ eV? Is it
caused by the end of the galactic component?\\
-- What is the relation between cut-off due to leakage out of the Galaxy and cut-off
effects due to maximum energies in sources?\\
A careful investigation of the region 10$^{16}$ -- 10$^{18}$ requires arrays with area 
$\sim$ 1 km$^2$ or more, but with with much smaller spacing than that of
arrays for ultra-high energies like AGASA, Yakutsk or AUGER.

\begin{figure}[h]
\begin{center}
\includegraphics*[width=0.6\textwidth, angle=0,clip]{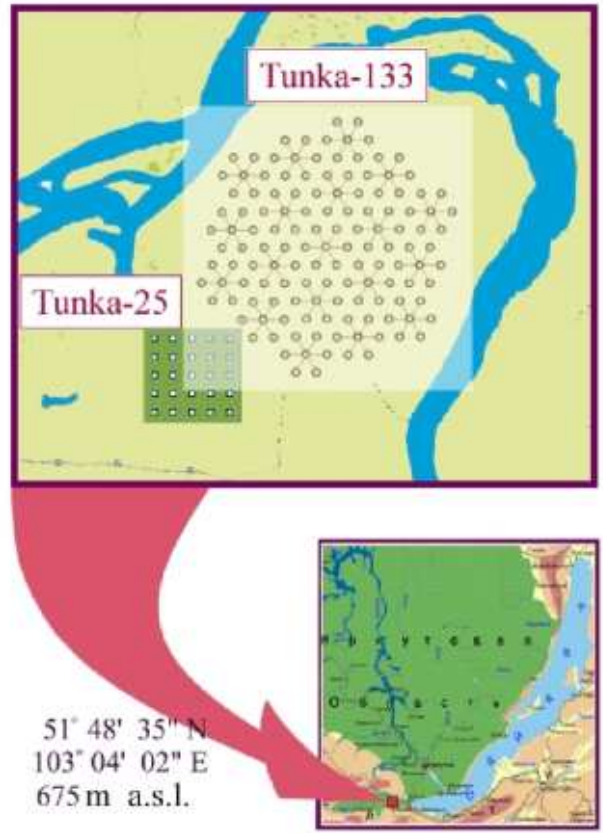}
\caption{\label{fig2} Location of Tunka-25 an the proposed Tunka-133 array.}
\end{center}
\end{figure}

\section {The Tunka-133 array}

The most reliable and accurate method of cosmic ray energy measurement in the
specified energy range is the method of EAS Cherenkov light observation. A series of such
experiments has been carried out in the Tunka Valley (50 km from Lake Baikal) since 
more than 10 years: Tunka-4 \cite{tunka-4}, Tunka-13 \cite{tunka-13} and Tunka-25
\cite{tunka-25}. The latter array consists of 25 detectors on the basis of
PMT QUASAR-370 and 4 detectors with pulse shape recording on the basis 
of PMT Torn-EMI D669. To study charged particles at the surface, an Auger-like
water Cherenkov detector was added to the array. The detectors cover an area
of $\sim$ 0.1 km$^2$. The new array Tunka-133 \cite{Tunka133} will consist of 133 optical
detectors on the basis of PMT EMI 9350 (diameter of photocathode of 20 cm).
The plan of the array is shown on fig.\ref{fig2}. The 133 detectors are grouped in 19 clusters,
each composed of seven detectors. The distance between detectors will be 85 m,
similar to Tunka-25. Due to this fact the accuracy of EAS parameter reconstruction
will be the same as for Tunka-25 \cite{Tunka}. The accuracy of the core location
is $\sim$6 m, that of energy determination $\sim$ 15\%. The accuracy of the X$_{max}$
determination (from the lateral disribution sharpness and pulse duration)
is $\sim$ 25 g/cm$^2$.

The construction of Tunka-133 will expand the area of the existing array with energy
threshold of $\sim$ 10$^{15}$eV by almost ten times. During one
year of operation (400 hours), Tunka-133 will record $\sim 5\cdot10^5$ events with
energy above $3\cdot10^{15}$eV, $\sim 300 $ events with energy higher than 10$^{17}$ eV
and a handful events with energy higher than 10$^{18}$ eV.

\section{Technical description}


The optical detector will be similar to that of Tunka-25. The angular
aperture is defined by the shadowing of the container. It is 20$^0$ for
100\% efficiency and reduces to 50\% at zenith angles $> 45^0$. Apart from
the PMT, the detector box contains the high voltage supply,  
the preamplifier, a light emitting diode for both amplitude and time
calibration and the control mechanics of the lid protecting the
PMT from sunlight and precipitation. To provide the necessary
dynamic range (10$^4$), two analog signals - one of them pre-amplified by
a factor 20, the other without amplification - will be transfered  to
the central electronics hut of each cluster.

\begin{figure}[h]
\begin{center}
\includegraphics*[width=0.9\textwidth,angle=0,clip]{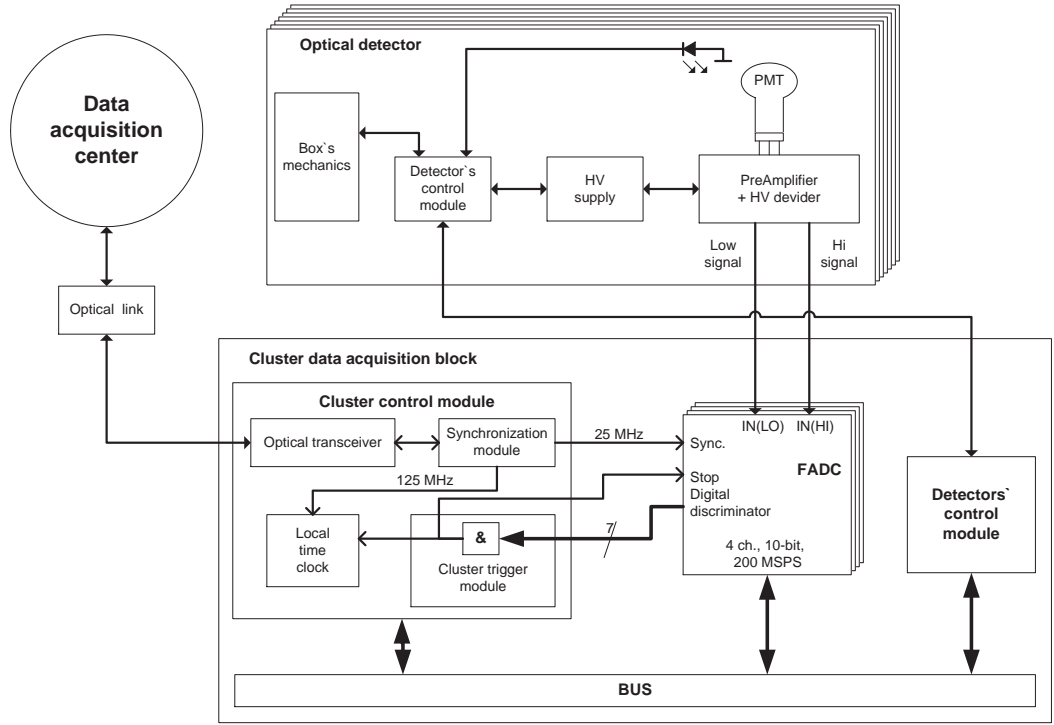}
\caption{\label {fig3} Functional scheme of cluster electronics}
\end{center}
\end{figure}

The functional scheme of the cluster electronics is shown in fig.\ref{fig3}.
 Each cluster is connected to the DAQ center through a multi-wire cable 
 containing four copper wires and four optical fibres.

The cluster data acquisition block contains optical transceiver, synchronization module,
local time clock and cluster trigger module. Optical transceivers operating at  
 1250 MHz are responsible on 
 data base transmition and forming of synchronization signals on the frequency 125 MHz
for cluster clock.
 A special procedure will be used to 
measure absolute time shifts between the clock in the DAQ center and the cluster clocks. 

The cluster trigger( the local trigger) is formed on the coincidence of pulses above
threshold in three optical detectors in a time window of 0.5 $\mu s$. The time of local trigger is
measured by cluster clock. 

Measuring channels are designed on the basis of 10 bit ADC AD9410 and XILINX microchips 
FPGAs. High-speed ADC with 5 ns step width are used for the pulse shape digizitation and
allow to discard traditional TDC approaches to determine the arrival time.
The ADC output data are writing to cyclic memory. The cluster trigger starts the readout 
256x10 bits (256x5ns $\approx$ 1.2 $\mu$s) from each ADC to the buffer memory of cluster control model.
A clock freguency 200 MHz is generated from the 25 MHz signals delivered form control model to all
measuring boards. 

\section{Schedule of the array deployment}

The cluster electronics with its software is prepared and tested. A test
of the first optical detectors and analog electronics in the Tunka valley will be
carried out in October 2005.

The first 4 clusters (Tunka-28) will be deployed in the fall
of 2006. The test operation of the array will extend over the winter
season 2006-2007.

The further installation of array depends on the financial support. In
case of sufficient support, the rest of array could be constructed in
2006 and 2007 and installed in Summer/Autumn 2007.

\section{Conclusion}

A 1-km$^2$ array in the Tunka valley (Buriatia,Siberia) is planned to record
EAS from cosmic rays of super-high energies by their
Cherenkov light. It will allow to study cosmic
rays by a single method covering uniformly the energy range 10$^{15}$ -- 10$^{18}$ eV.
This includes the classical knee 
at $\sim 3\cdot 10^{15}$ eV as well as the second knee at a few
$10^{17}$ eV, allowing to study features of
the spectrum probably connected with the 
transition from galactic to extra-galactic cosmic rays. 
With this installation we are going to study up to which maximal energy particles
in supernova remnants are accelerated and will also 
provide a low-energy calibration to much larger installations like AUGER.

\section{Acknowledgements}
The present work is supported by the Russian Ministry of Education and Science, by
the Russian  Fund of Basic Research 
(grants 03-02-16660, 05-02-04010, 05-02-16136) and by the Deutsche
Forschungsgemeinschaft DFG (436 RUS 113/827/0-1).


\begin{thebibliography}{99}

\bibitem{EASTOP}
M.~Aglietta et al., Astropart.Phys. 20(2004) 641

\bibitem{Kaskade}
 T.~Antoni et al., astro-ph/0505413 (2005)
 
\bibitem{Space}
M.~Ahres et al. Astropart.Phys. 21(2004) 565
 

\bibitem{Tunka}
D.~Chernov et al., astro-ph/0411139 (2004)

\bibitem{tunka-4}
S.~V.~Bryanski et al.//Proc. 24th ICRC, Roma,2,724,1995.

\bibitem{tunka-13}
O.~A.~Gress et al.,//Nucl.Phys.B(Proc.Suppl.) 75A.229,1999.

\bibitem{tunka-25}
N.~M.~Budnev et al.,//Proc.27th ICRC, Hamburg, 1, 581, 2001.

\bibitem{Tunka133}
D.~Chernov et al., astrop-ph/0411218 (2004)


\end{thebibliography}
\end{document}